\documentclass[a4paper,12pt]{article}
\usepackage{amssymb}
\usepackage{epsf}
\usepackage{cite}

\def\preprint#1#2{\noindent\hbox{#1}\hfill\hbox{#2}\vskip 10pt}
\begin {document}
\begin{titlepage}

\preprint{ITP-UH-18/99, SINP/TNP/99-31}{October 1999}
\vfill

\begin{center}
{\Large\sc
 Phase diagram of an exactly solvable\\ $t$--$J$ ladder model }

\vfill

{\sc Holger Frahm}
\vspace{1.0em}

\emph{
Institut f\"ur Theoretische Physik, Universit\"at Hannover\\
  D-30167 Hannover, Germany}
\vspace{1.0em}

and
\vspace{1.0em}

{\sc Anjan Kundu}\footnote{email: anjan@tnp.saha.ernet.in}
\vspace{1.0em}

\emph{
Saha Institute of Nuclear Physics,  
	 Theory Group,
	 1/AF Bidhan Nagar\\ Calcutta 700 064, India}
\end{center}

\vfill

\begin{quote}
We study a system of one-dimensional $t$--$J$ models coupled to a
ladder system.  A special choice of the interaction between
neighbouring rungs leads to an integrable model with supersymmetry,
which is broken by the presence of rung interactions.  We analyze the
spectrum of low-lying excitations and ground state phase diagram at
zero temperature.
\end{quote}

PACS-Nos.\ {71.10.Pm, 
	71.10.Hf, 
        75.10.Lp 
     }

\vspace*{\fill}
\setcounter{footnote}{0}
\end{titlepage}

Systems of one-dimensional magnetic or electronic chains coupled to
so-called ladders have attracted increasing attention as a scenario in
which two-dimensional physics can be studied by starting from well
established one-dimensional theories \cite{DaRS92,BDRS93,gors:94}.
The properties of the undoped systems, i.e.\ spin ladders, have been
thoroughly investigated by a variety of methods
\cite{DaRS92,BDRS93,gors:94,brmn:96,bkmn:98,komi:98r} and recently
integrable models with tunable interaction parameters have been
introduced \cite{PZ,FrRo97,Fei99,Wang99,MuTa99,BaMa99a,BaMa99}.
%
To understand the properties of the doped ladder systems the dynamics
of the holes created under doping in the background of
antiferromagnetically interacting spins has to be studied
\cite{Bose95}.  This question has been studied within a model of
coupled $t$--$J$ chains forming a ladder-like structure in a series of
papers \cite{Bose94,Bose95,Bose99} with interesting results.  Within
this approach, however, multi hole scattering processes or the
dynamics of anti-bonding hole states could not be accessed.

Our aim here is to propose a modified $t$--$J$ ladder model, which
becomes integrable at certain limits and therefore allows us to
investigate the properties of this system exactly and in more detail.
  In
the thermodynamic limit, we can analyze further the nature of the
ground state and predict the onset of excitations, thereby identifying
the phase diagram of the system.
The system we consider  consists of two coupled $t$--$J$ chains
labeled $a=1,2$.  The electrons with spin $\sigma$ on site $j$ of the
chains are described in terms of canonical Fermi operators
$c^{(a)}_{j\sigma}$.  In terms of these the chain Hamiltonian is
$H_{t-J}^{(a)}=\sum_j \left\{H_{t-J}^{(a)}\right\}_{jj+1}$ with
\begin{equation}
  \left\{H_{t-J}^{(a)}\right\}_{jk}=   
  -t{\cal P}\left(\sum_{\sigma=\uparrow\downarrow}
	c^{(a)\dagger}_{\sigma j} c^{(a)}_{\sigma k} +h.c.
   \right){\cal P}+\
  J\left({\bf S}^{(a)}_j{\bf S}^{(a)}_{k} 
  -{1 \over 4} {n}^{(a)}_j{n}^{(a)}_{k}\right)
  + {n}^{(a)}_j+{n}^{(a)}_{k}
\label{Htj}
\end{equation}
where ${\cal P}$ projects out double occupancies, ${\bf S}^{(a)}_j$
are spin operators on site $j$ and $n^{(a)}_j = n^{(a)}_{j\uparrow} +
n^{(a)}_{j{\downarrow}}$ is the total number of electrons on site $j$.

The Hamiltonian of the ladder model we consider then reads
\begin{equation}
  H= \sum_a H_{t-J}^{(a)} + H_{\rm int} + H_{\rm rung} - \mu \hat{N}\ .
\label{H0}
\end{equation}
%
Here, $\mu$ is the chemical potential coupling to the total number of
electrons in the system, $H_{\rm int}$ is the interaction between
neighbouring rungs of the ladder, given by the expression
%
$H_{\rm int}=-\sum_j  \left\{H_{t-J}^{(1)}\right\}_{jj+1} 
		  \left\{H_{t-J}^{(2)}\right\}_{jj+1}$,
and $H_{\rm rung}$ contains all the interactions present on a single rung
of the ladder and reads
\begin{equation}
  H_{\rm rung} =  \sum_j \left\{
	-t'{\cal P}\left(\sum_{\sigma=\uparrow\downarrow}
	c^{(1)\dagger}_{\sigma j} c^{(2)}_{\sigma j} +h.c.
   \right){\cal P}
	+J'\left({\bf S}^{(1)}_j{\bf S}^{(2)}_{j} 
	  -{1\over4} {n}^{(1)}_j{n}^{(2)}_{j}\right)
	+ V {n}^{(1)}_j{n}^{(2)}_{j} \right\}\ .
\label{Hrung}
\end{equation}
This is similar to the terms in (\ref{Htj}) describing the interaction
of electrons on the bonds within a rung but contains independent
coupling constants $t'$, $J'$ and an additional Coulomb interaction of
strength $V$.  Below we shall use a basis of states on a single rung
which are eigenstates to (\ref{Hrung}): at half filling (one electron
per site) the electrons can form a spin singlet $|s\rangle \equiv {1
\over \sqrt 2} (|\uparrow \downarrow\rangle - |\downarrow\uparrow
\rangle)$ or triplet state $|t_+\rangle \equiv |\uparrow\uparrow\rangle$
etc.\ ($|\sigma\tau\rangle=c^{(1)\dagger}_\sigma c^{(2)\dagger}_\tau
|00\rangle$).
Doping of the system creates the states $|\sigma_\pm\rangle \equiv{1
\over \sqrt 2} (|\sigma 0\rangle\pm |0 \sigma\rangle)$, with
$\sigma=\uparrow $ or $\downarrow$, describing a single electron with
spin $\sigma$ in a bonding (anti-bonding) state or the Fock vacuum
(two-hole state) $|d\rangle \equiv |00\rangle$ on a rung.
Below we shall be particularly interested in the strong coupling
regime $J'\gg1$, $V\gg\mu+|t'|$ near half filling.  This implies that
the triplet states $|t_\alpha\rangle$ in the two-particle sector will
be energetically unfavourable.

To study the phase diagram of the system at low temperatures we note
that by excluding the triplet states and choosing the coupling
constants on the legs of the ladder as $J=2t=2$ the Hamiltonian
(\ref{H0}) can be rewritten as 
\begin{equation}
  H=-\sum_j \Pi_{jj+1} - \sum_{l=1}^5 A_l N_l + {\rm const.}
\label{H1}
\end{equation}
where $N_l$, $l=1,2$ ($3,4$) is the number of bonding (anti-bonding)
single particle rung states with spin $\uparrow,\downarrow$ and $N_5$
is the number of empty rungs (on the remaining $N_0=L-\sum_l N_l$
rungs the two electrons form a singlet).  The (graded) permutation
operator $\Pi_{jk}$ interchanges the states on rungs $j$ and $k$
giving a minus sign if both rungs are fermionic, i.e.\ singly
occupied.  In the restricted Hilbert space we are considering here we
can absorb $J'$ into the definition of $V$ which allows to express the
potentials $A_l$ in terms of the coupling constants in (\ref{Hrung})
as
\begin{eqnarray}
&&	A_1=A_2\equiv \mu_+ =t'-\mu+V,\ 
\nonumber\\
&&	A_3=A_4\equiv \mu_- =-t'-\mu+V,\
	A_5 \equiv \tilde V=-2\mu+V 
\label{param}
\end{eqnarray}
The relative strength of these potentials will determine the ground
state and spectrum of low lying excitations in the ladder.

Note that the Hamiltonian (\ref{H1}) is that of a $gl(2|4)$ invariant
super spin chain in the presence of external symmetry breaking fields
and the grading of the states $|s\rangle$ and $|d\rangle$ is bosonic
while the two doublets $|\sigma_\pm\rangle$ carry fermionic grading.
This model is integrable by means of the Bethe Ansatz (BA) (see e.g.\
\cite{Kulish85,schl:92c,eks:94a,schl:rev}).
In the strong coupling regime near half filling, where the ground
state shows little deviations from the dimerized state $\otimes_{j}
|s\rangle_j$ doping will lead to the condensation of either the
fermionic single hole states $|\sigma_\pm\rangle$ or the bosonic
double hole state $|d\rangle$ into the ground state -- depending on
the relative strength of the chemical potentials $\mu_-$, $\mu_+$ and
the rung interaction $\tilde V$.  For simplicity we shall
consider the case $\mu_+=\mu_-\equiv\tilde\mu$ below, i.e.\ two
degenerate single particle bands.  Due to the grading different sets
of BA equations have to be considered to describe these processes (see
e.g. \cite{Kulish85,esko:92,foka:93}):
since the Bethe Ansatz states are highest weight in the $gl(2|4)$
algebra we have to deal separately with two regions in the low
temperature phase diagram of the $t$--$J$ ladder (\ref{H0}) between
half- and quarter-filling, corresponding to $4N_5 < \sum_{l=1}^4 N_l$
and $4N_5 > \sum_{l=1}^4 N_l$, respectively.

For the first case we have to order the rung basis as $|s\rangle$,
$|\sigma_\pm\rangle$, $|d\rangle$ and the eigenstates of (\ref{H1})
are parameterized by solutions of the BA equations corresponding to
this  grading (BFFFFB)
\begin{eqnarray}
   \left[e_1(\lambda_j^{(1)})\right]^L &=&
     \prod_{k=1}^{M_2} e_1(\lambda_j^{(1)}-\lambda_k^{(2)})\ ,
\nonumber\\
   \prod_{k\ne j}^{M_r} e_2(\lambda_j^{(r)}-\lambda_k^{(r)}) &=&
     \prod_{l=1}^{M_{r-1}} e_1(\lambda_j^{(r)}-\lambda_l^{(r-1)}) 
     \prod_{m=1}^{M_{r+1}} e_1(\lambda_j^{(r)}-\lambda_m^{(r+1)})\ ,
   \quad r=2,\ldots,4\ ,
\label{baeBF}\\
   \prod_{k=1}^{M_4} e_1(\lambda_j^{(5)}-\lambda_k^{(4)}) &=& 1\ 
\nonumber
\end{eqnarray}
with $e_n(x) \equiv (x+in/2)/(x-in/2)$. Here the number of roots on the
$r$th level is related to the numbers $N_l$ in (\ref{H1}) by $M_{r+1}
= M_{r}- N_r$ (we define $M_0\equiv L$).
The energy of the corresponding state is
\begin{equation}
  E\left(\{\lambda_j^{(r)}\}\right) = -2 M_1
	+ \sum_{j=1}^{M_1} {1\over(\lambda_j^{(1)})^2+1/4}
        - \sum_{l=1}^5 A_l N_l\ .
\label{EE}
\end{equation}
In the thermodynamic limit $L\to\infty$ with $M_r/L$ fixed the
solutions of (\ref{baeBF}) are built on two types of so-called strings,
namely complexes
\[
 \lambda^{(r-n)}_k = \xi^{(r)} + \frac{i}{2}(n-2k)\ ,\
    k=0,\ldots,n\ ,\
    0\le n<r\ {\rm and}\ 2\le r\le4
\]
corresponding to bound states of holes with different spin and parity
with densities $\rho^{(r)}(\xi)$ and
\[
 \lambda^{(r)}_{k} = \Lambda^{(r)}_{m} + \frac{i}{2}(m+1-2k)\ ,\
    k=1,\ldots,m\ ,\
    2\le r \le4
\]
corresponding to bound spin states with densities
$\sigma_m^{(r)}(\Lambda)$, $r=2,3,4$. On level $1$ and $5$ all
solutions are real with densities $\sigma_1^{(r)}(\lambda^{(r)})$,
$r=1,5$.

By minimizing the free energy of the system (see e.g.\
\cite{schl:92c,schl:rev}) one finds that at most the bound states of
single particle rung states $|\sigma_\pm\rangle$ with density
$\rho^{(4)}$ and empty rungs $|d\rangle$ with density $\sigma_1^{(5)}$
are present in the ground state.  The densities are determined through
the Bethe Ansatz integral equations
\begin{eqnarray}
  \rho^{(4)}(x) &=&  a_4(x)
	- \int_{s}{\rm d}y\ K(x-y) \rho^{(4)}(y)
	- \int_{d}{\rm d}y\ a_1(x-y) \sigma_1^{(5)}(y)\ ,
\nonumber\\
\label{dens1}\\
  \sigma_1^{(5)}(x) &=& \int_{s}{\rm d}y\ a_1(x-y) \rho^{(4)}(y)\ 
\nonumber
\end{eqnarray}
where $K(x) = \sum_{n=1}^3a_{2n}(x)$ and $a_n(x) = 2n/\left(\pi ({
4x^2 + {n}^2 })\right)$.
The corresponding dispersions $\epsilon_s$ and $\epsilon_d$ are given
by similar integral equations
\begin{eqnarray}
  \epsilon_s(x) &=& 2\pi a_4(x) -4(2+\tilde\mu)
	- \int_{s}{\rm d}y\ K(x-y) \epsilon_s(y)
	+ \int_{d}{\rm d}y\ a_1(x-y) \epsilon_d(y)\ ,
\nonumber\\
\label{dressE1}\\
  \epsilon_d(x) &=& \tilde\mu - \tilde V
	- \int_{s}{\rm d}y\ a_1(x-y) \epsilon_s(y)\ .
\nonumber
\end{eqnarray}
The chemical potential and the rung interaction determine the ranges of
integration 
\[
  \int_\alpha{\rm d}y\ =\left\{\int_{-\infty}^{-Q_\alpha}
		    +\int_{Q_\alpha}^\infty\right\}{\rm d}y\ 
\]
through the condition $\epsilon_\alpha(\pm Q_\alpha)=0$.  In terms of
the solution of (\ref{dens1}) the density of holes is 
\begin{equation}
   n_h = 2\int_s{\rm d}y\ \rho^{(4)}(y) 
       - \int_d{\rm d}y\ \sigma_1^{(5)}(y)\ .
\end{equation}
For $\tilde V<\tilde\mu<-2$ both $\epsilon_s$ and $\epsilon_d$ are
positive and hence the dimer state is the ground state of the system.
For larger chemical potential single hole states $|\sigma_\pm\rangle$
with dispersion $\epsilon_s$ begin to condensate into this ground
state.  The electronic properties in this phase are those of the degenerate
supersymmetric $t$--$J$ model \cite{schl:92c}.  For $\tilde
V>\tilde\mu -\int_s{\rm d}y\ a_1(y) \epsilon_s(y)$ the system enters a
phase where in addition double hole rung states $|d\rangle$ with energy
$\epsilon_d$ will be present in the ground state.

For $\tilde\mu<\tilde V$ we have to order the the rung states as
$|s\rangle$, $|d\rangle$, $|\sigma_\pm\rangle$ basis leading to a
different set of Bethe Ansatz equations corresponding to BBFFFF
grading ($M_6\equiv 0$).
\begin{eqnarray}
   \left[e_1(\lambda_j^{(1)})\right]^L &=&
     \prod_{k\ne j}^{M_1} e_2(\lambda_j^{(1)}-\lambda_k^{(1)})
     \prod_{k=1}^{M_2} e_1(\lambda_k^{(2)}-\lambda_j^{(1)})\ ,
\nonumber\\
     \prod_{l=1}^{M_{1}} e_1(\lambda_j^{(2)}-\lambda_l^{(1)}) 
	&=& 
     \prod_{m=1}^{M_{3}} e_1(\lambda_j^{(2)}-\lambda_m^{(3)})\ ,
\label{baeBB}\\
   \prod_{k\ne j}^{M_r} e_2(\lambda_j^{(r)}-\lambda_k^{(r)}) &=&
     \prod_{l=1}^{M_{r-1}} e_1(\lambda_j^{(r)}-\lambda_l^{(r-1)}) 
     \prod_{m=1}^{M_{r+1}} e_1(\lambda_j^{(r)}-\lambda_m^{(r+1)})\ .
   \quad r=3,\ldots,5\ .
\nonumber
\end{eqnarray}
(Note that due to the reordering of the basis the number of double hole
rung states is now $N_1$.)  The energy of the corresponding Bethe
Ansatz eigenstates is again given by (\ref{EE}) with $A_1=\tilde V$ and
$A_l=\tilde\mu$ for $2\le l\le5$.
Similarly as for (\ref{baeBF}) the solutions of these equations can be
classified in strings with densities $\sigma_m^{(r)}$ for positive
integers $m$ and $r=1,3,4,5$ and single hole bound states with
densities $\rho^{(r)}$, $r=3,4,5$.  Only real roots with density
$\sigma_1^{(2)}$ are possible on the second level of (\ref{baeBB}).
The ground state configuration is completely determined by
\begin{eqnarray}
  \rho^{(5)}(x) &=& - \int_{s}{\rm d}y\ K(x-y)\rho^{(5)}(y)
	+ \int_{d}{\rm d}y\ a_4(x-y) \sigma_1^{(1)}(y)\ ,
\nonumber\\
\label{dens2}\\
  \sigma_1^{(1)}(x) &=&  a_1(x)
	+ \int_{s}{\rm d}y\ a_4(x-y) \rho^{(5)}(y)
	- \int_{d}{\rm d}y\ a_2(x-y) \sigma_1^{(1)}(y)\ .
\nonumber
\end{eqnarray}
The dispersion of these excitations are determined through the
integral equations
\begin{eqnarray}
  \epsilon_s(x) &=& 4(\tilde V-\tilde\mu)
	- \int_{s}{\rm d}y\ K(x-y) \epsilon_s(y)
	+ \int_{d}{\rm d}y\ a_4(x-y) \epsilon_d(y)\ ,
\nonumber\\
\label{dressE2}\\
  \epsilon_d(x) &=& 2\pi a_1(x)-(2+\tilde V)
	+ \int_{s}{\rm d}y\ a_4(x-y) \epsilon_s(y)
	- \int_{d}{\rm d}y\ a_2(x-y) \epsilon_d(y)\ .
\nonumber
\end{eqnarray}
and the density of holes is $n_h = \int_d{\rm d}y\ \sigma_1^{(1)}(y)\
- 2\int_s{\rm d}y\ \rho^{(5)}(y)$.

With these equations the phase diagram of the $t$-$J$ ladder
(\ref{H1}) can be completed: for $\tilde\mu<\tilde V<-2$ the system is
in the dimer phase, increasing the rung interaction $\tilde V$ the
system enters a phase where the ground state is a Fermi sea of double
hole states with dispersion $\epsilon_d$ moving in the background of
dimer states $|s\rangle$.  Increasing $\tilde\mu$ in this phase beyond
$\frac{1}{8} \left(7\tilde V -2\right)$ both double and single hole
rung states are present in the ground state configuration.

\begin{figure}[ht]
\begin{center}
\leavevmode
\epsfxsize=0.65\textwidth
\epsfbox{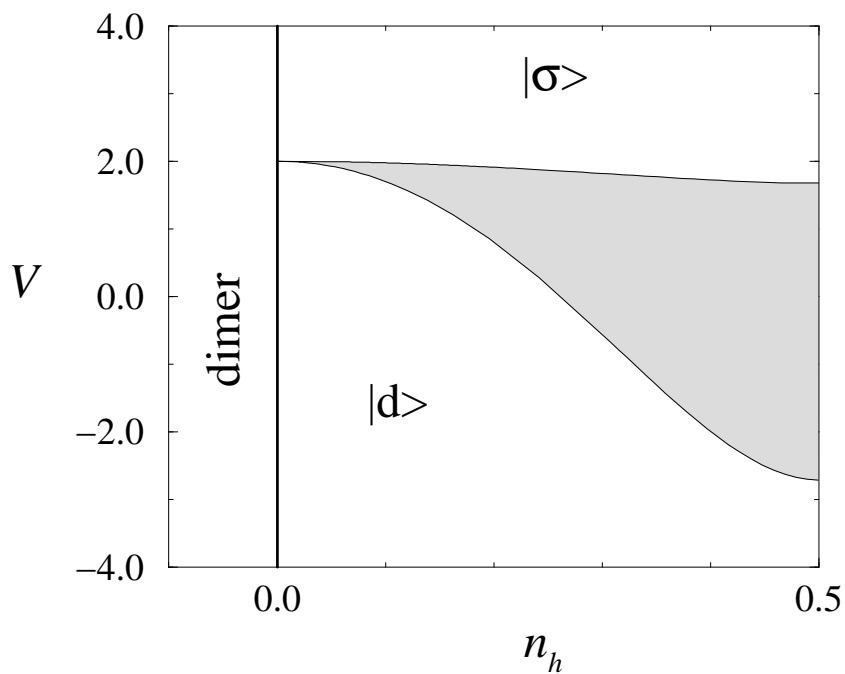}
\end{center}
\caption{Phase diagram of the $t$--$J$ ladder with rung interaction
(\protect{\ref{Hrung}}) for $t'=0$ and $J'\gg1$ as a function of the
hole concentration $n_h$: For large repulsive $V$ the ground state is
a Fermi sea of single hole states $|\sigma_\pm\rangle$ propagating in
the background of rung dimers.  For sufficiently strong attractive
rung interaction the double hole rung states $|d\rangle$ are
energetically favourable.  In the intermediate region (shaded) dimers
\emph{and} both types of hole rung states coexist in the ground state
configuration.
\label{fig:phases}}
\end{figure}

Numerical integration of the integral equations for the densities and
dressed energies allows to determine the phase boundaries as a
function of the hole density $n_h$ and the  rung
interaction $V=2\tilde\mu-\tilde V$ in Eq.~(\ref{Hrung}).  In
Fig.~\ref{fig:phases} the resulting phase diagram is shown.  For
$V\gtrsim 2$ condensation of the single hole states
$|\sigma_\pm\rangle$ is energetically favourable.  For sufficiently
strong \emph{attractive} rung interaction, however, doping of the
system first introduces double hole rung states $|d\rangle$.
Interestingly, we find that for
$2>V>2-\frac{1}{2}\left(\psi\left(\frac{7}{8}\right) -
\psi\left(\frac{7}{8}\right)\right)\approx 1.676$ ($\psi(x)$ is the
digamma function) increasing the hole concentration can drive the
system from the latter phase into the one without double hole rung
states.

In summary we have studied the phase diagram of a system of coupled
$t$-$J$ models in the framework of an integrable system.  By choosing
properly the interaction between the neighbouring rungs 
 the resulting ladder model can be brought into the form of a
$gl(2|4)$-invariant exchange model in the presence of symmetry
breaking fields, which enter the Hamiltonian as kinetic and interaction
terms on a single rung.  The spectrum of excitations and zero
temperature phase diagram have been determined by means of the Bethe
Ansatz.  Extensions are possible by assigning different energies to
the single particle rung states of different parity ($t'\ne0$) and
including the triplet rung states ($J'$ finite).  As has been shown
for the undoped case new phases can be expected for finite $J'$
\cite{Wang99}.

HF acknowledges partial support by the Deutsche Forschungsgemeinschaft
under Grant No.\ Fr~737/2 and AK acknowledges with thanks 
the Fellowship Grant from the Alexander von Humboldt
Foundation.


\setlength{\baselineskip}{14pt}

\end{document}